\documentstyle[preprint,aps,epsf]{revtex}

\begin{document}
\title {Patterns of consumption in socio-economic  models
with heterogeneous interacting agents.}

\author{Giulia  Iori$^{(a)}$ and Vassilis Koulovassilopoulos$^{(b)}$ } 
\address{ (a) Dept.  of Accounting, Finance and Management, \\
University of Essex  Wivenhoe Park, Colchester CO4 3SQ, UK  \\
{\small  Email: iorig@essex.ac.uk}\\
(b)  Dept. of Nuclear and Particle Physics, \\
University of Athens, Panepistimiopoli, Ilisia 15771, Athens, Greece \\
{\small Email: vkoulov@cc.uoa.gr}
}

\maketitle

\begin{abstract}
We study  consumption behaviour in systems  with
heterogeneous interacting agents. 
Two different models are introduced, respectively with  long and  short range 
interactions among agents.
At any time step an agent decides
whether or not to consume a good, doing so if this  provides positive
utility. Utility is affected by  idiosyncratic preferences and costs
as well as externalities from other agents. Agents are ranked in classes 
 and recognize peer, distinction and aspiration groups. We simulate the
system  for different choices  of the parameters and identify different
complex patterns: a steady state regime  with a variety of
consumption modes of behaviour, and a wave/cycle regime.
The cases of fad and value goods are both analyzed.

\end{abstract}
\newpage
\section{Introduction}

A great body of literature
has been devoted to the effects that  interactions  among consumers or
firms  (henceforth called agents)  have on the macroeconomic variables
of  social-economic systems (Aoki 1996).  
Applications range from   consumption theory, economic  growth, opinion
formation, stock market prices and emergence of de facto standards in
technology choice (see Kirman (1997) and Durlauf (1997) for recent reviews).

A common  feature of many aggregate variables
is that they exhibit oscillatory behaviour, showing boom-and-bust patterns. 
Examples are  fad and bandwagon  behaviour in sociology, business cycles
in economics,  bubbles in stock market prices, wave behaviour in the
adoption of innovation technology.  Boom-and-bust cycles have been
observed (Persons et al. 1996)  also  in the adoption of financial technologies,
 like the CMO (collateralized mortgage obligations) and the 
LBO (leveraged-buyout) waves of 1980s, the merger and aquisition waves 
of the 1890s, 1920s and 1960s as well the takeover wave of the 1980s. 

Such  oscillatory behaviour can be explained in terms of
interacting agents if an individual's consumption decision depends
 on the behaviour of different   group of consumers.
Various  models have been introduced (Kirman 1996, Durlauf 1996), with
interactions among agents depending on the agents' ``distance'' (such
as, for example, differences in wealth) but typically in these 
 models agents interact in a symmetric way
with each other. 
 
Recently Cowan, Cowan and and Swan (1998), hereafter (CCS), introduced a model where
the utility of an individual agent can be positively or negatively
affected  by the choices of  different groups of agents and  consumption
is driven by  peering, imitiation and distinction effects.
In CCS (1998), consumers are ordered  according to their social status
and are affected by  the behaviour of other agents  depending on  their
relative  location on the spectrum. Agents wish 
to distinguish themselves from those who are below
and emulate their peers and those who are above in the social spectrum.

Even though this model is capable of reproducing consumption waves,
which emerge from the   interplay between aspiration  and distinction
effects, it lacks  an important ingredient, namely costs of consumption. 

We reformulate here the model of paper CCS (1998) in the  framework
of statistical mechanics of disordered
systems, incorporating   costs of
consumption which vary across agents and are exogenously determined.  
The dynamical properties of the model are explored, by numerical
simulations, for different  choices of the parameters. Various
complex patterns are found.

\section{Model} 

We consider N agents,
ordered on a one-dimensional continuum space according to their
``wealth''  $w_i$. In this paper wealth serves as an index of social
status rather than the source of a budget constraint, as discussed below.
 A more realistic situation with consumers arranged over a
multidimensional space (accounting, for example, for differences in  age,
education, etc.)  should be considered, but for simplicity
we will only discuss here the case $d=1$.
Agents' wealth is chosen randomly, it is  uniformly distributed between zero
and $W_0$, and does not change with time.

Time evolves discretely and at
each step the state of the agent $i$ is characterezied by a variable
$S_i$ which can take only two values $\pm 1$; if agent $i$ chooses to
consume then $S_i=1$ while  if (s)he chooses not to consume then $S_i = 
-1$.   

At time zero a new product appears in the market and each  agent decides
whether or not to consume one indivisible unit of it, doing so if this  provides positive
utility (for example a new restaurant opens at time zero and in each
subseqent time  period  agents decide
whether to visit it or not). Utility $Y_i(t)$ \footnote{It corresponds to a Hamiltonian
interaction: $H=\sum_{i=1}^{N} Y_i(t) S_i(t)$.} 
is affected by  idiosyncratic preferences and costs
as well as externalities from other agents: 

\begin{equation}
Y_i(t) = \frac{1}{N} \sum^N_{j=1} J_{ij} S_j(t) + G(w_i) + C  c_i + N  n_i(t)
\label{Model} 
\end{equation}

Individual    costs, $c_i$ are uniformly  distributed in the interval
$(-1,0)$ and the functional form of the intrinsic value $G(w_i)$ of the
good for different agents  will be specified below.
In this partial equilibrium set-up with only one good, the decision on whether 
to buy the good or not depends only on whether the ``marginal'' utility  
of buying one unit exceeds the ``marginal'' cost. In this sense, all
agents are assumed to possess sufficient liquidity at all points of
time and wealth constraints are never binding.

A product with $G = 0$  is called a ``fashion'' good,
while a ``status'' good has a positive intrinsic value  that might be
well-suited to the characteristics or tastes of a particular class of
consumers. 
The  idiosyncratic noise term $n_i$, describes individual preferences
which can change with time, and are random uniformly distributed in the
interval  $(-1, 1)$.  

\begin{figure}[htb]
  \epsfxsize 13cm  \centerline{\epsffile{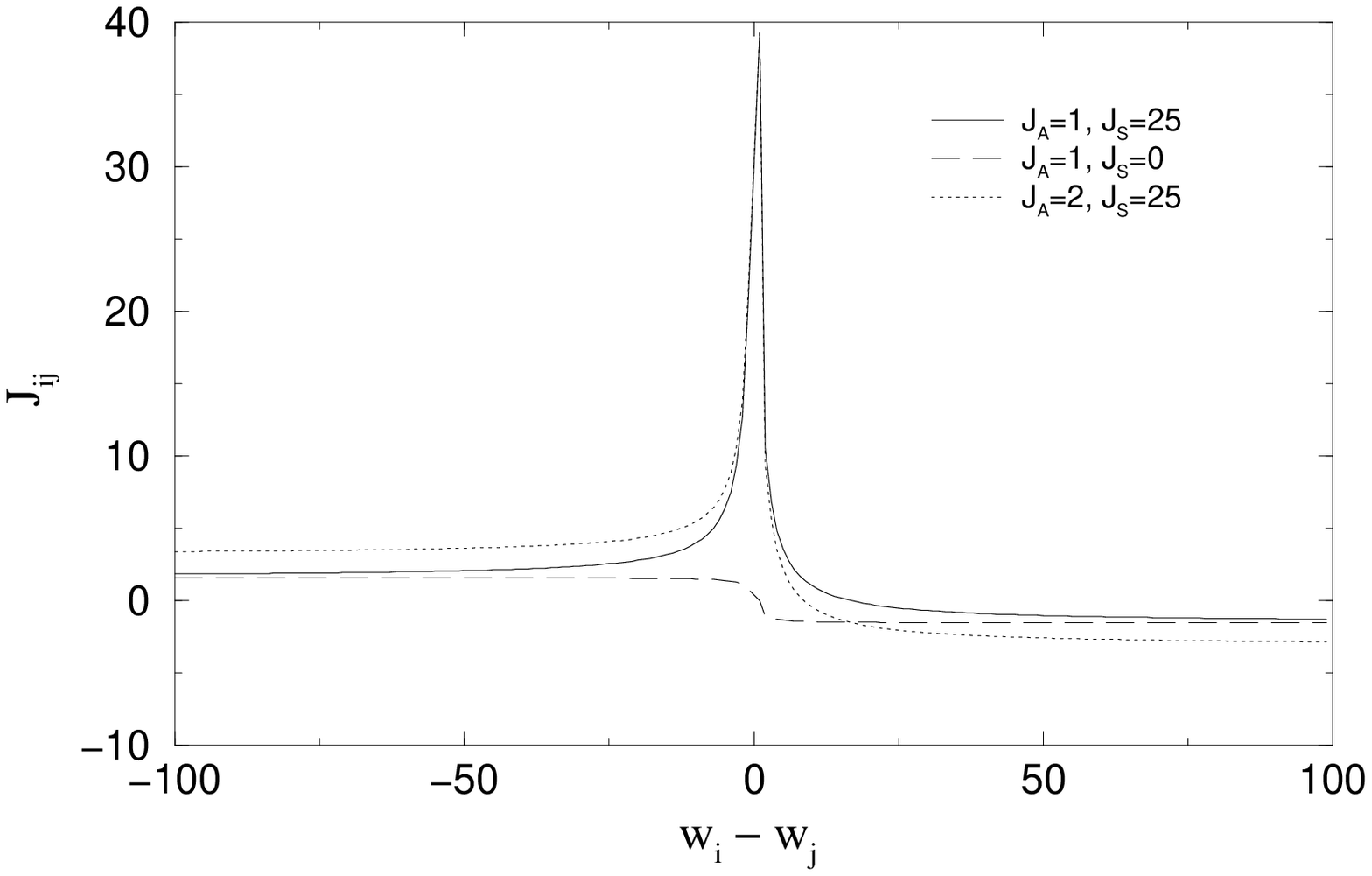}}
\vskip .5cm
  \epsfxsize 13cm \centerline{\epsffile{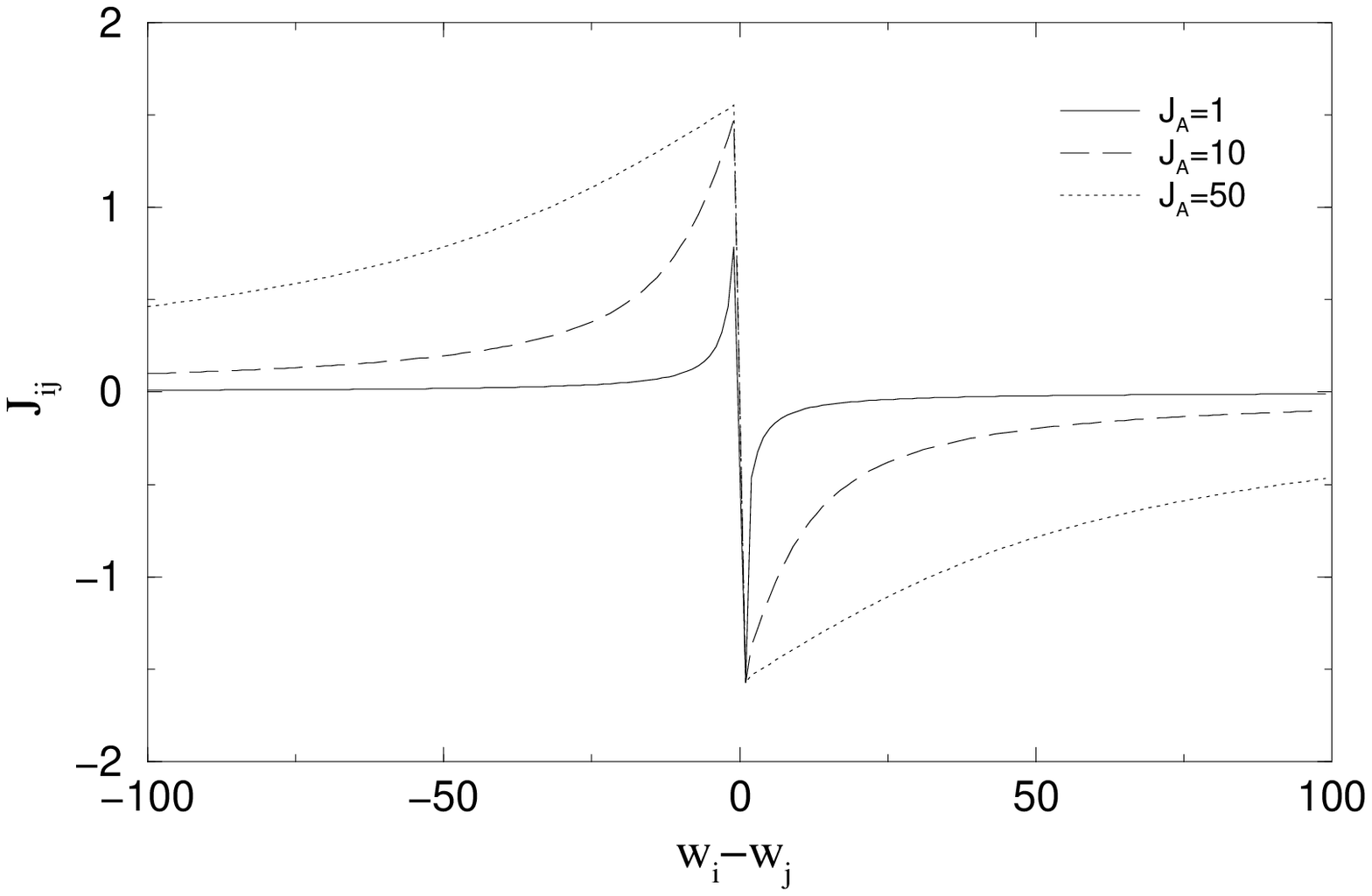}}
\vskip .5cm
\caption{ $J_{ij}$ respectively for Model I (top) and Model II
(bottom). Notice that while in Model I the $J_{ij}$ saturate to a finite
value at large distances (long-range interactions), in Model II the $J_{ij}$
are short range and decay quickly to zero.} 
 \label{Pot}
\end{figure}

Each  agent interacts with all the others,
and the coupling constants $J_{ij}$ are functions of the agents' status
according to:
\begin{equation}
J_{ij} = -J_A \arctan (w_i -w_j) +
 J_S \left[ \frac{\pi}{2} - \arctan |w_i -w_j| \right]  \label{Jij}
\end{equation}
The coefficients $J_A, J_S$ are taken positive. 
The asymmetric term, proportional to $J_A$,
gives a negative contribution to the utility function $Y_i$ if $w_i>w_j$ and a 
positive contribution if $w_i<w_j$. This means that agent $i$ wishes to
distinguish  herself from the poorer  while imitating the
richer.
The second contribution, proportional to $J_S$, always generates positive
utility and expresses peering effects among consumers of similar status.
Both of these contributions saturate with distance. In
Fig.~(\ref{Pot}.a) $J_{ij}$ is plotted as a function of $(w_i - w_j)$. 

At each time $t$, we let  agents re-evaluate their consumption
decision, relative to the one taken one step before,  on the basis of 
the utility $Y_i(t)$ they receive 
at the {\it same} time. This amounts to assuming that agents are myopic
or that they face a complete absence of opportunities for intertemporal 
arbitrage, {\it i.e.}~ there are no capital markets. Agents choose
simultaneously, that is, our system evolves with parallel dynamics.
 We have further allowed the possibility that only a 
fraction $\alpha$ of the total population reconsider their
decision, with this subpopulation picked randomly at each time step.
In the following though we shall consider only the case $\alpha=1$.

An alternative scenario that will be considered (which we 
call Model II henceforth)  is when $J_S=0$ and the asymmetric
influences exercised by agents on the  others  are given by: 
\begin{equation}
J_{ij} = \arctan\frac{J_A}{w_i-w_j}
\end{equation}
$J_{ij}$ is shown in Fig.~(\ref{Pot}.b).
In this case there are only  distinction-aspiration effects, which are
stronger for agents of similar status and vanish at large distance. 
This describes the situation where agents segregate in social-economic groups
so that among members  of the same group there is a tendency to outshine
their peers and to  ignore the rest.

In both cases, the coupling constants $J_{ij}$ are random numbers
(positive or negative)  which provides an analogy with spin-glass models
(Mezard et al. 1987). The peculiar properties of spin-glasses have attracted
considerable  attention (Aoki 1996, Arthur et al. 1988, Arthur et al.1997) in   social-economic
disciplines, particularly towards the possible implications, in that
context, of the  degeneracy of the  equilibrium state and of the  slow
relaxation dynamics to equilibrium (aging).

We stress though that in our formulation the coupling constants are not
symmetric and we do not expect our model to share the  nice properties of spin
glasses (Crisanti et al. 1988, Cugliandolo et al. 1997).
Nonetheless a study (Iori et al. 1997) on the Sherrington-Kirkpatrick (SK)
 model with a non-hamiltonian contribution 
(i.e. taking $J_{ij} = J^S_{ij} + \epsilon J^A_{ij}$) 
has  shown  that  both the dynamical behaviour  and the nature of
equilibrium  (at least for small lattices) survives for a small
asymmetric perturbation. In view of this result it could be interesting
to  repeat the study of Iori et al. (1997) when $\epsilon $ is zero for all
but a certain fraction of the $(i,j)$ pairs, for which pairs, $\epsilon$
can be large. This case would be similar to the model proposed in
eq.~(\ref{Model}). 

Eventually, the case of our model in eq.~(\ref{Model}) with $J_A = 0$ 
(i.e. considering only peering effects) would be  a natural framework to study
the problem of  technological progress.
In this case, indeed, while peering  finds a natural interpretation in
terms of the advantage that interacting  firms have (for example those
who share a partnership) for being  at a similar technological 
level  (see Arenas et al.(1999), aspiration and distinction effects
do not affect the payoff  a firm receives   from  adopting  a
new technology.
When $J_A = 0$, eq.~(\ref{Model}) describes a modified Random Field
Ising Model (RFIM) with site dependent couplings.
 At zero temperature, when adiabatically driven by an external field,
the RFIM is characterized by  avalanche dynamics. 
In our model the role of the external field is played by costs.

\section{Simulations and Results}
\subsection{Model I}

We have performed computer simulations to study the model in
eq.~(\ref{Model}) above for  different choices of the parameters  and
explored the various  patterns observed. We studied both the time
evolution of  total consumption as well as the shape of the 
distribution of consumption  across the social spectrum. 

The pictures  we show refer to a relatively small lattice of $V=800$
consumers (for the graphics we are limited by the size of the screen). 
We analyzed  however larger lattices  ($V=4000$) and results do not 
qualitatively change with the lattice size.
Agents' wealth is uniformly distributed in the interval $(0, W_0)$
with $W_0=100$ and does not change with time.

We assume that, when a new good enters the market, agents initially do not
consume it ($S_i(0)=-1$).  We could have  alternatively chosen a 
nonhomogeneous initial condition  (with consumption distributed across
agents with a certain probability)  and study how the system 
evolves towards the equilibrium state.
 
\begin{figure}[htb]
  \epsfxsize 10cm \centerline{\epsffile{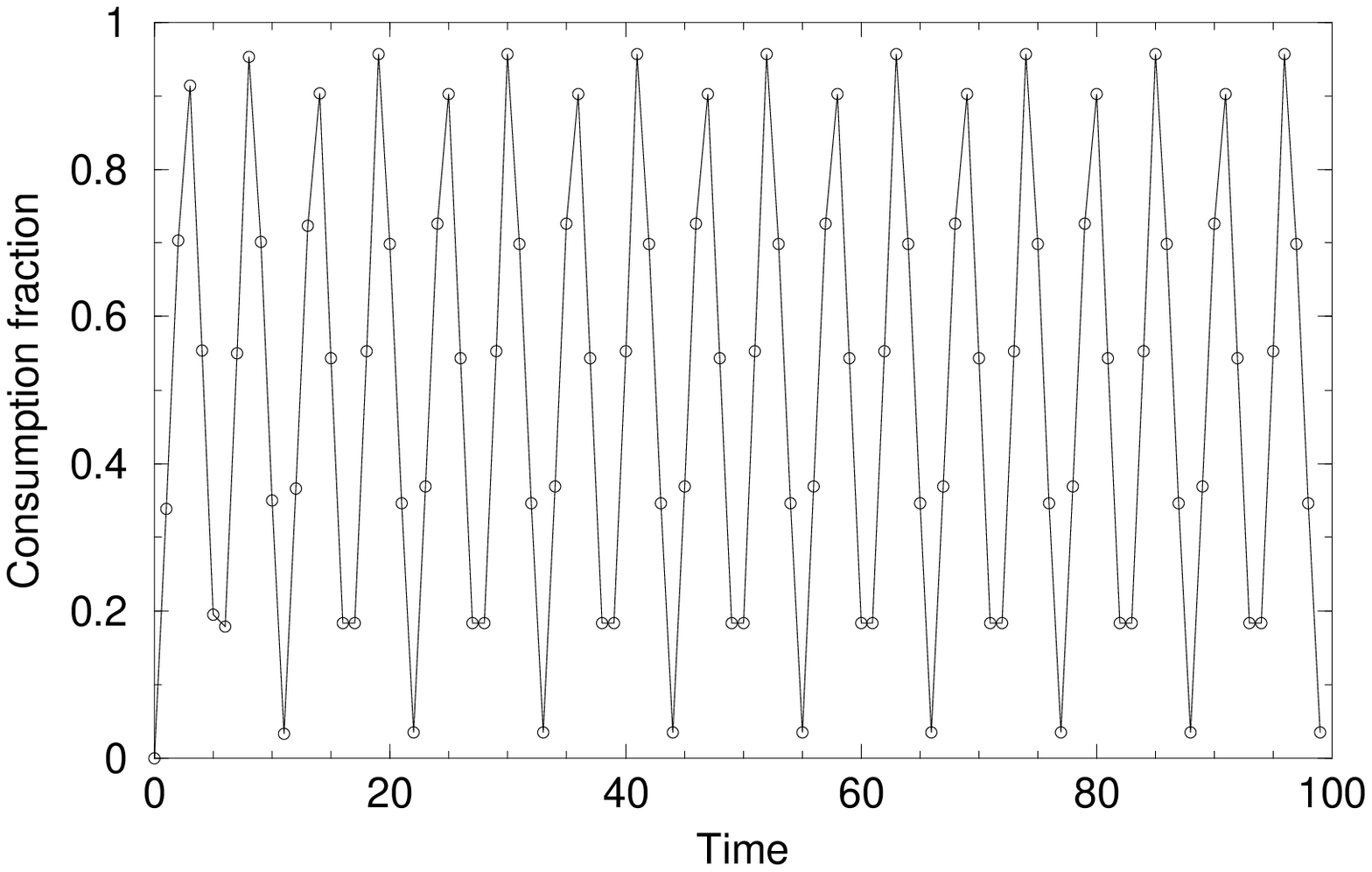}}
  \epsfxsize 10cm  \centerline{\epsffile{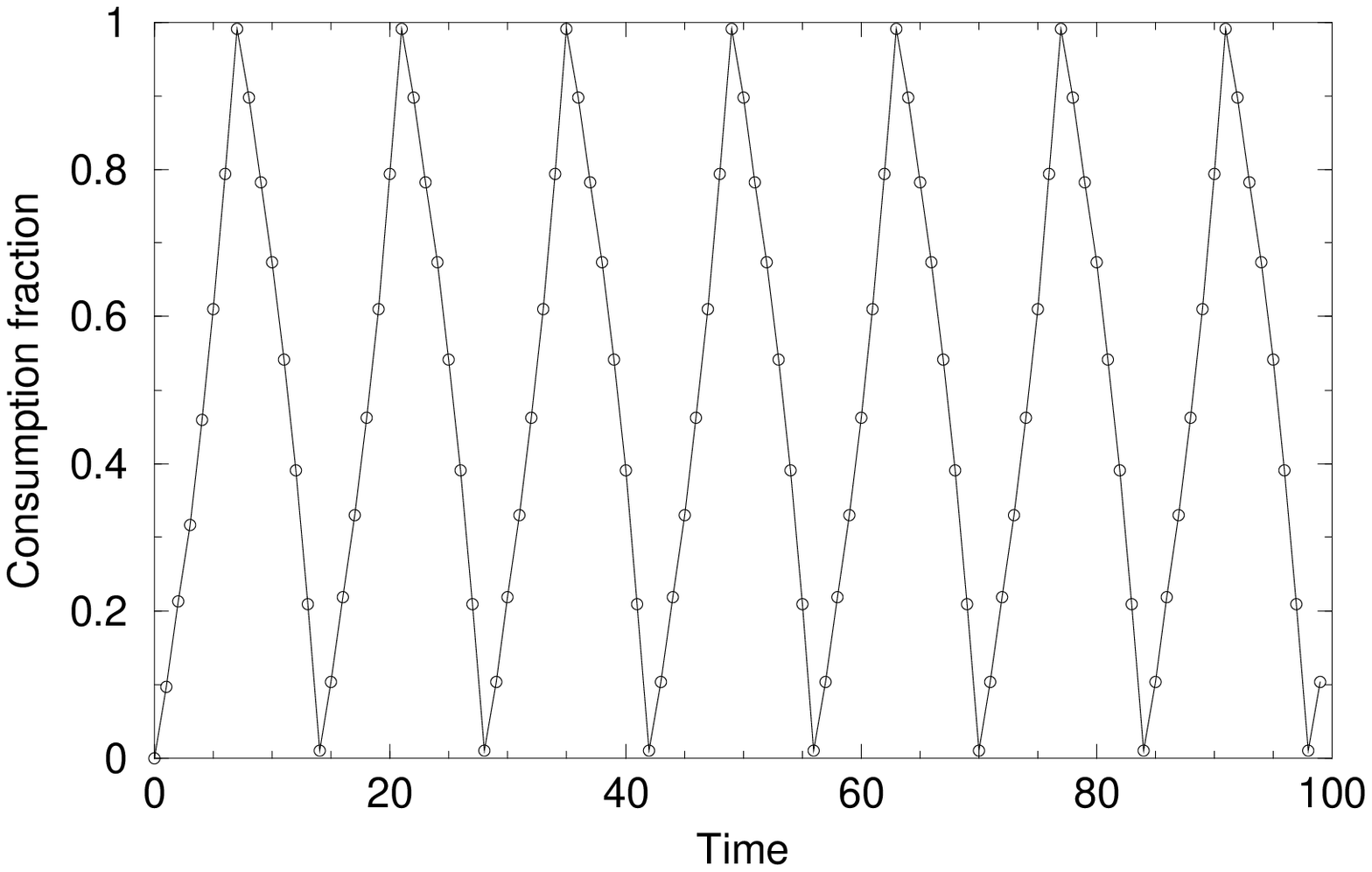}}
  \epsfxsize 10cm \centerline{\epsffile{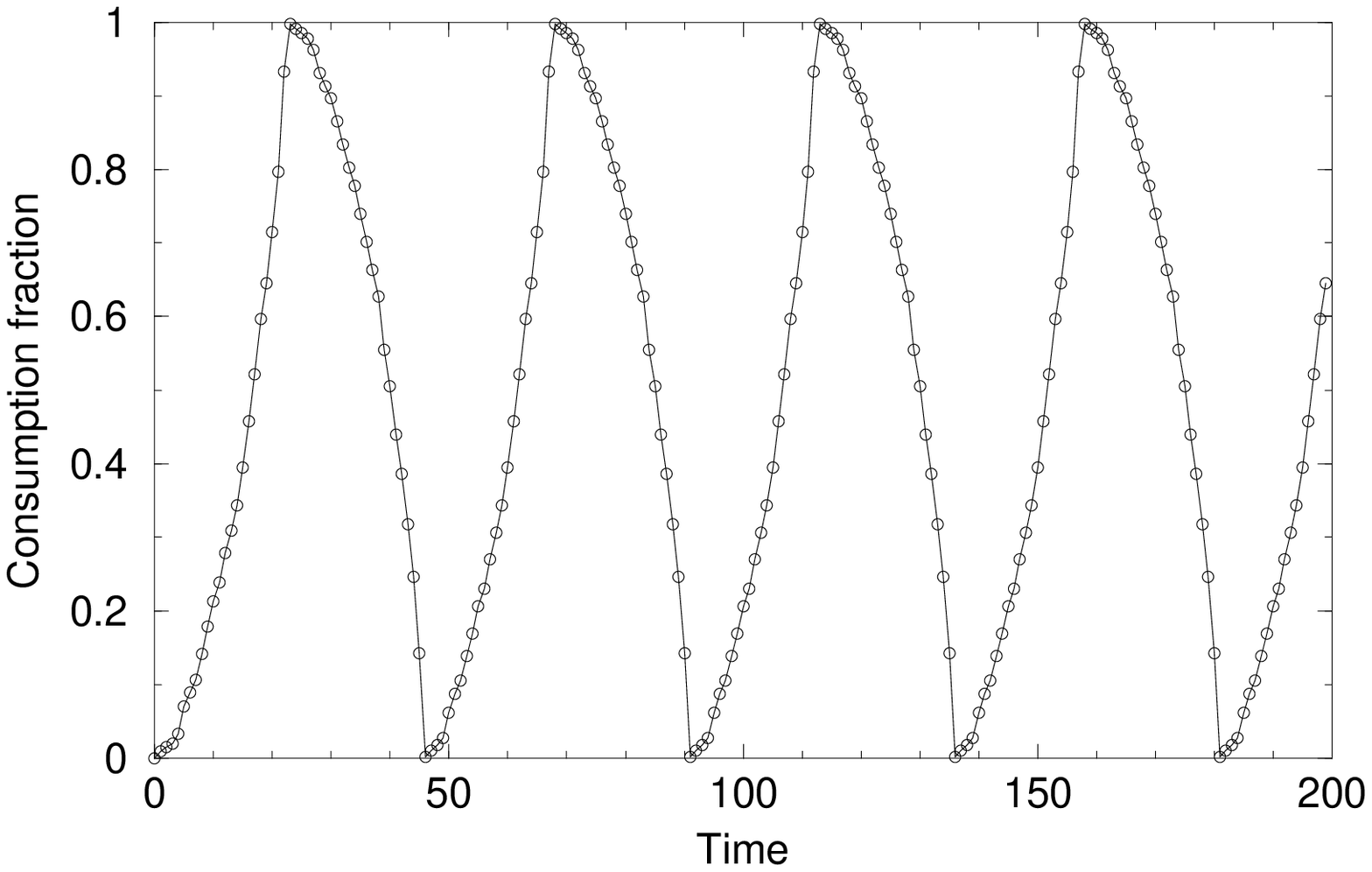}}
\vskip .5cm
\caption{Total consumption as a function of time in Model I. $J_A=1$ and three
different values of $J_S=5, 15, 25$ are considered respectively from top
to bottom. Here $C =G = N = 0$.} 
 \label{waves}
\end{figure}

\begin{figure}[htb]
  \epsfxsize 14cm \centerline{\epsffile{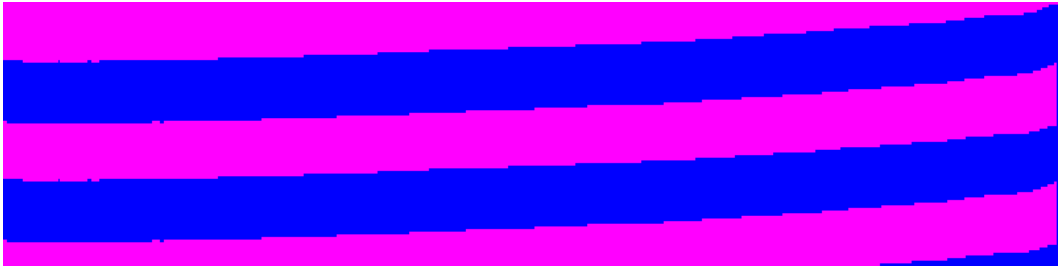}}
\vskip .5cm
\caption{Wave dynamics in Model I, with the dark color for those agents
consuming and light color for those not consuming. Wealth is increasing
from left to right along the horizontal axis. The vertical axis
represents time which increases from top to bottom. Here $C=G=N=0, J_A=1,
J_S=25$.}   
 \label{waves_time}
\end{figure}

\begin{figure}[htb]
  \epsfxsize 14cm \centerline{\epsffile{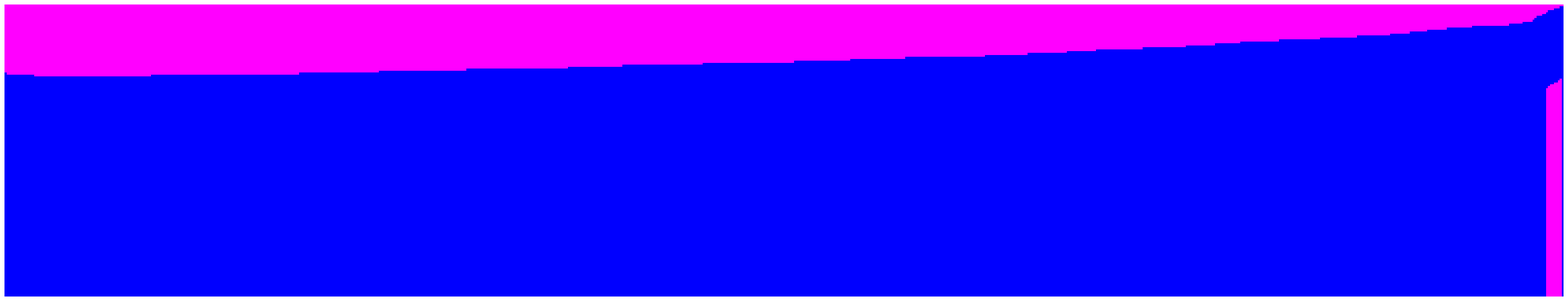}}
\vskip 1cm
  \epsfxsize 14cm \centerline{\epsffile{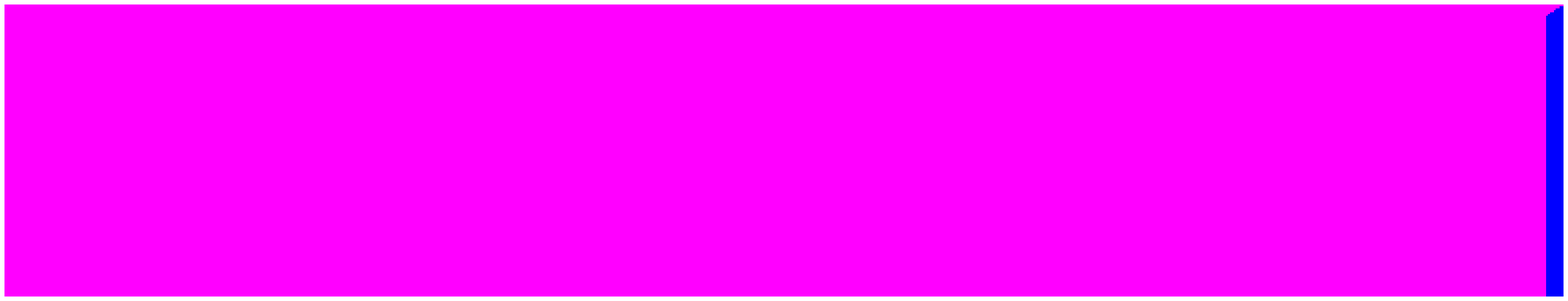}}
\vskip .5cm
\caption{Steady-state behaviour for Model I in region II at $J_A=1, J_S=29.4$ (top),
and region III at $J_A=1, J_S=30.5$ (bottom), where $C=G=N=0$.}  
 \label{steady_states}
\end{figure}

\begin{figure}[htb]
  \epsfxsize 13cm \centerline{\epsffile{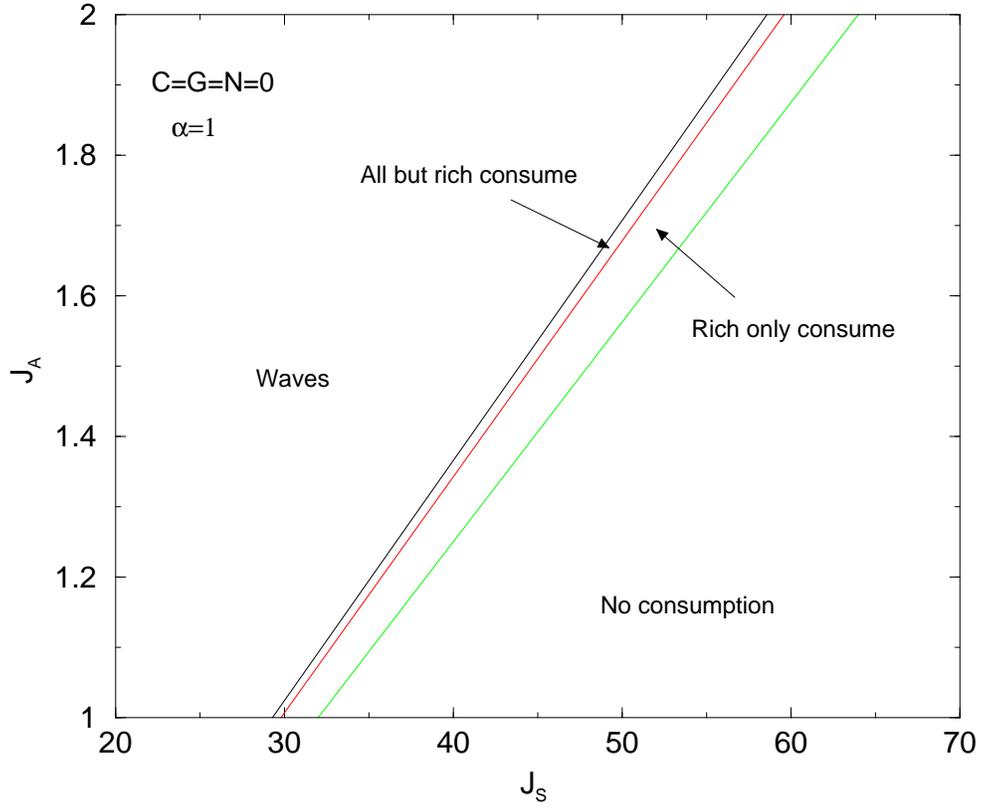}}
\caption{Phase diagram in $J_S-J_A$ space for $C=G=N=0, \alpha=1$.}  
 \label{phase_J1_J2_N0.00}
\end{figure}

Initially we study  the case of a fashion good ($G = 0$), choosing 
 $C = N = 0$. We explore the different consumption patterns
 when  fixing $J_A$ and increasing $J_S$. For $J_S$ small (region I) the
 total  consumption exhibits waves,
with amplitude less than $V$ and high frequency. As $J_S$ is increased
the period of oscillations increases and the amplitude becomes almost
equal to $V$. The situation is depicted in Fig.~(\ref{waves}) for
$J_A=1$ and $J_S=2, 15, 25$. 
In Fig.~(\ref{waves_time}) we show  a typical pattern,  for $J_A=1, J_S=
 25$, as a function of time (time increases from top to bottom).  
Notice that it is the rich who start consuming first in order to
distinguish themselves from the poor. At  the beginning of
each cycle consumption propagates rather slow but it spreads faster as it
moves down to the poorer. 
As we keep increasing $J_S$ though, the waves disappear and the system
develops a steady-state behaviour with a variety of consumption
patterns: above the 
value $J^c_S$ where waves disappear, we first find a narrow region
(region II) where all but the richest consume. Still increasing $J_S$
we find a region (region III) where only the rich consume but not the
rest, while if we further increase $J_S$ (region IV) nobody consumes as
distinction effects are not strong enough to initiate the
process. Simulations for these three regions are shown in
 Fig.~(\ref{steady_states}). 
The precise location of these four regions in the $J_A, J_S$ region is
depicted in Fig.~(\ref{phase_J1_J2_N0.00}) for a $800$ lattice.

\begin{figure}[htb]
  \epsfxsize 13cm
  \centerline{\epsffile{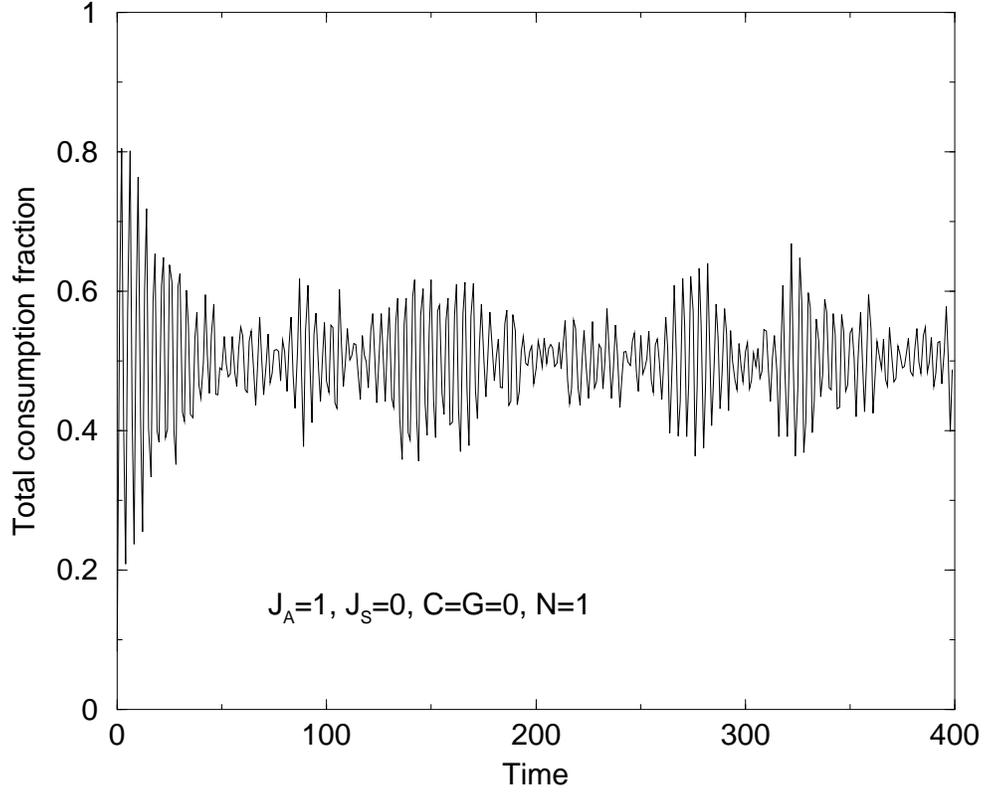}}
\caption{The idiosyncratic noise produces a modulated wave, in the
absense of costs, gains and peering effects, where the amplitude of
consumption changes periodically with time.}
\label{modulated}
\end{figure}

Up to now we have assumed that agents do not have specific preferences
for the new good. When introducing preferences, these act like the 
temperature in a physical system and can  induce a transition from an
equilibirum state to another. 
In our model, for example, as we turn on idiosyncratic noise ($N > 0$),
in region I waves  become blurred with consumption  greater than zero
and less than $V$, while  for $J_A=1, J_S=0, N=1$,
Fig.~(\ref{modulated}), we see curiously the emergence of modulated waves. 
Eventually, for  values of $N>0$, consumption waves  emerge also in
region IV (a larger $N$ is required the deeper one is in region IV)
where, at $N = 0$, nobody would consume.

\begin{figure}[htb]
\epsfxsize 13cm \centerline{\epsffile{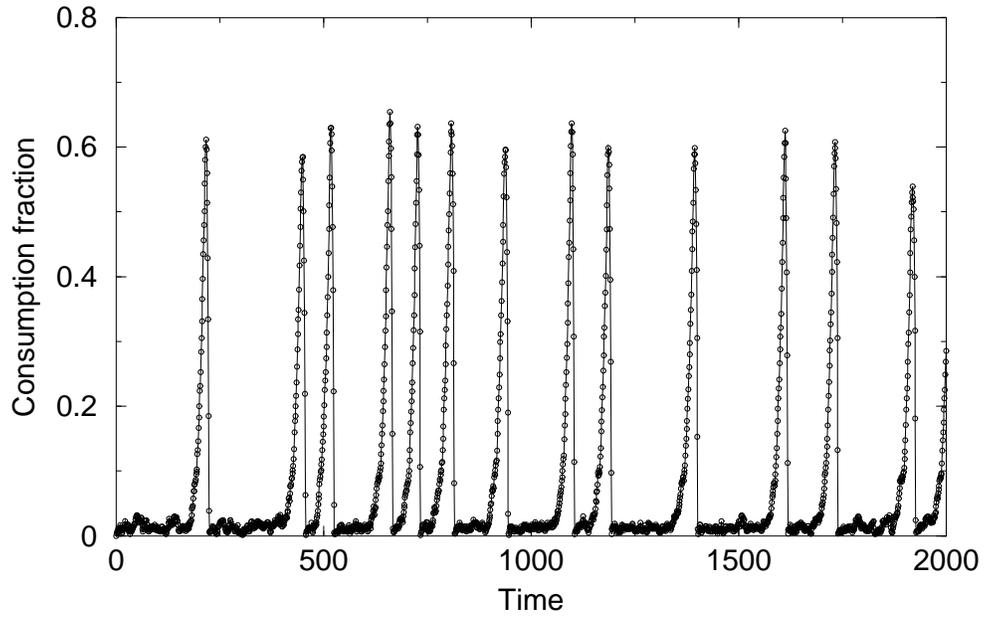}}
\caption{Waves in consumption with $J_A=1, J_S=25, C=1.29, G=0, N=0.9$
starting from $S_i(0)=-1$.}
 \label{FAD_Noise}
\end{figure}

Next we introduce costs ($C > 0$) while keeping $G = 0$ (or small compared to
costs). For cost below  a certain value (we find $C_c  = 0.465$
for $J_A = 1$ and $J_S = 25$) waves still emerge and propagate across
all the social spectrum, but if $C > C_c$  
the equilibrium state is characterized by nobody consuming.
Adding noise, at  $N \sim 1$
consumption waves, of varying amplitude, emerge spontaneously even at
$C > C_c$,  though at irregular time intervals  (see Fig.~(\ref{FAD_Noise})).
Nonetheless if  $N_i$ becomes  much  larger, waves disappear  and
total consumption, disorderely distributed,  fluctuates around a
constant level.  

\begin{figure}[htb]
  \epsfxsize 14cm \centerline{\epsffile{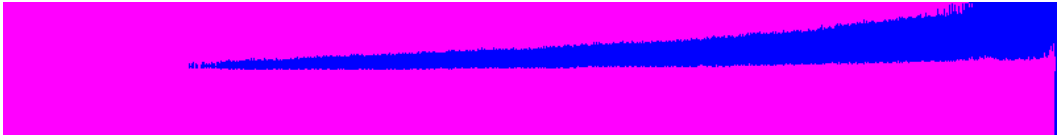}}
\vskip 1cm
\caption{If an initial condition is imposed that the richest  10\% are
consuming, and $N=G=0, C=1.29$, $J_A=1, J_S=25$ then a wave emerges
once. Time is $T=100$ steps.}   
 \label{initial_condition}
\end{figure}

Also initial conditions can have an important  role on the dynamical
selection of the different asymptotic states: 
in Fig.~(\ref{initial_condition}) we show that imposing an initial
condition with a fraction (here the top 10\% of the population) of the
agents consuming, then a consumption wave is generated even for $C > C_c$.

\begin{figure}[htb]
  \epsfxsize 15cm  \centerline{\epsffile{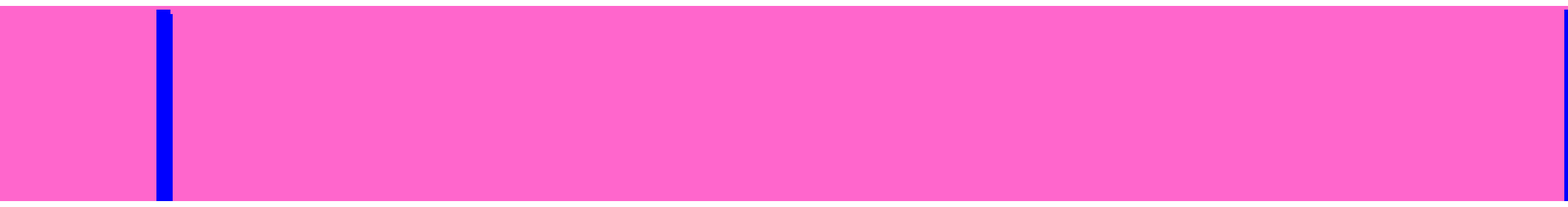}}
\vskip 1cm
  \epsfxsize 15cm  \centerline{\epsffile{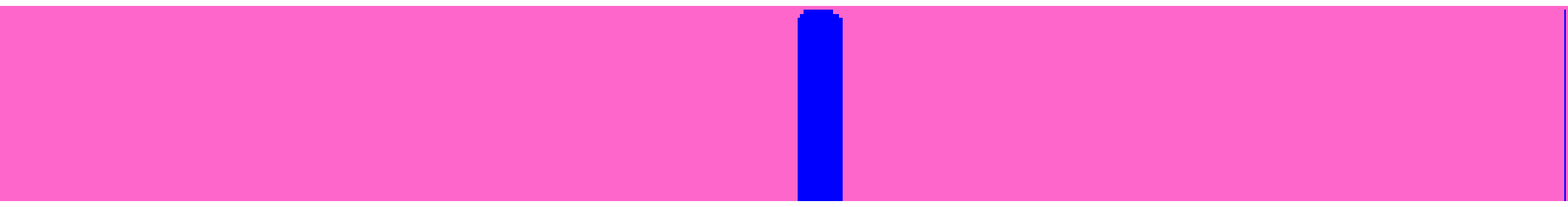}}
\vskip 1cm
  \epsfxsize 15cm  \centerline{\epsffile{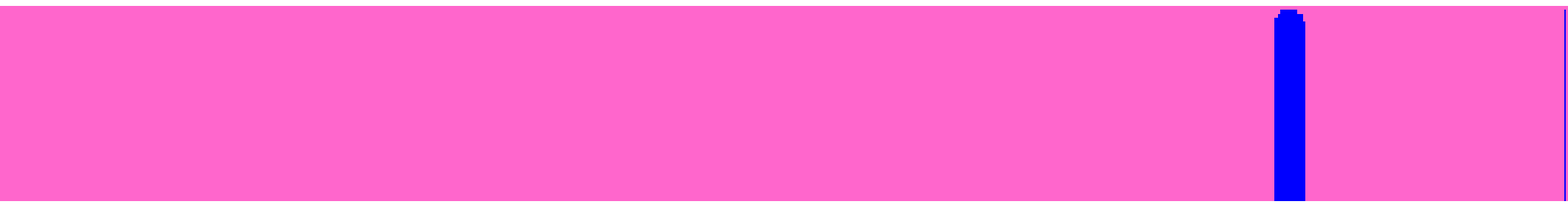}}
\vskip 1cm
  \epsfxsize 15cm  \centerline{\epsffile{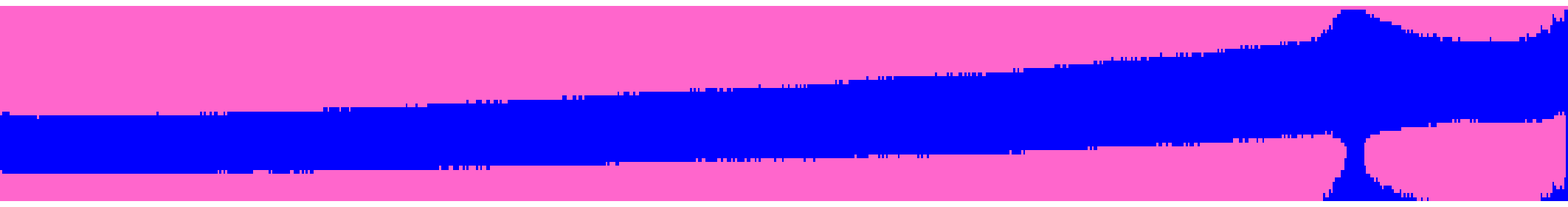}}
\vskip .5cm
\caption{Consumption behavior when $J_A=1$, $J_S=25$, $C=0.5$, $G=1$, $N=0$ and
$w_m = 10, 50, 80, 85$ (from top to bottom). Only when the good is suitable
to the consumers located  at the top of the social scale a consumption wave
propagates. In all other case the good enters and finds a stable niche.}
 \label{Niche}
\end{figure}

Finally  we consider the general case of a status good, which has an
intrinsic value ($G(w_i) > 0$) for some agents $i$ and its consumption entails
costs ($C_i > 0$). We consider the case of a  good which is more
suitable to this group of consumers whose wealth is distributed around a
given  value $w_m$, and we choose: 
\begin{equation}
G(w_i) = \frac{G}{(w_i - w_m)^2}
\end{equation}
The results which follow refer to the case  $J_A = 1, J_S = 25, G = 1, N = 0$.
We fix costs at a level higher than $C_c$, estimated before for the
case  $G = 0$ (below this value waves propagate even in the case of  a fashion
good). 
Different behaviours are found, depending on costs and/or $w_m$. For
$w_m < 85$, (we remind that agents'  wealth is distributed between zero and
$W_0 = 100$),   the good enters  the social
spectrum around $w_m$, possibly migrates through the closest social classes
and then finds a stable niche (see top  three cases in Fig.~(\ref{Niche})). 
If $w_m \sim W_0$,  for costs not too high (but still larger than
$C_c$),  waves emerge and  spread throughout the whole social spectrum 
(bottom case in  Fig.~(\ref{Niche})).

The effect of adding idiosyncratic preferences is again to induce waves  in
 regions where, in its absence, consumption would reach a constant
level and would be limited to a few consumers.

\subsection{Model II}

In this case, as we said, only aspiration/distinction effects affect the
consumption behaviour of agents. We will focus only on short range
interactions (e. g. choosing $J_A = 1$) while,  for larger values of 
$J_A$, as can be seen from Fig.~(\ref{Pot}a)),  the   potential becomes
longer range.
 
 We will  compare  Model II with Model I assuming  $J_S = 0$. While 
consumers feel in both cases  (only) distinction and aspiration
effects, their strength of interaction  changes  differently with
distance,  becoming zero at large distances for Model II, while it is
finite and  non-zero for Model I. Moreover at short distances the
potential is large and sharp for model II while it is  almost  zero for
Model I.

\begin{figure}[htb]
 \centerline{\epsfxsize 6.5cm \epsffile{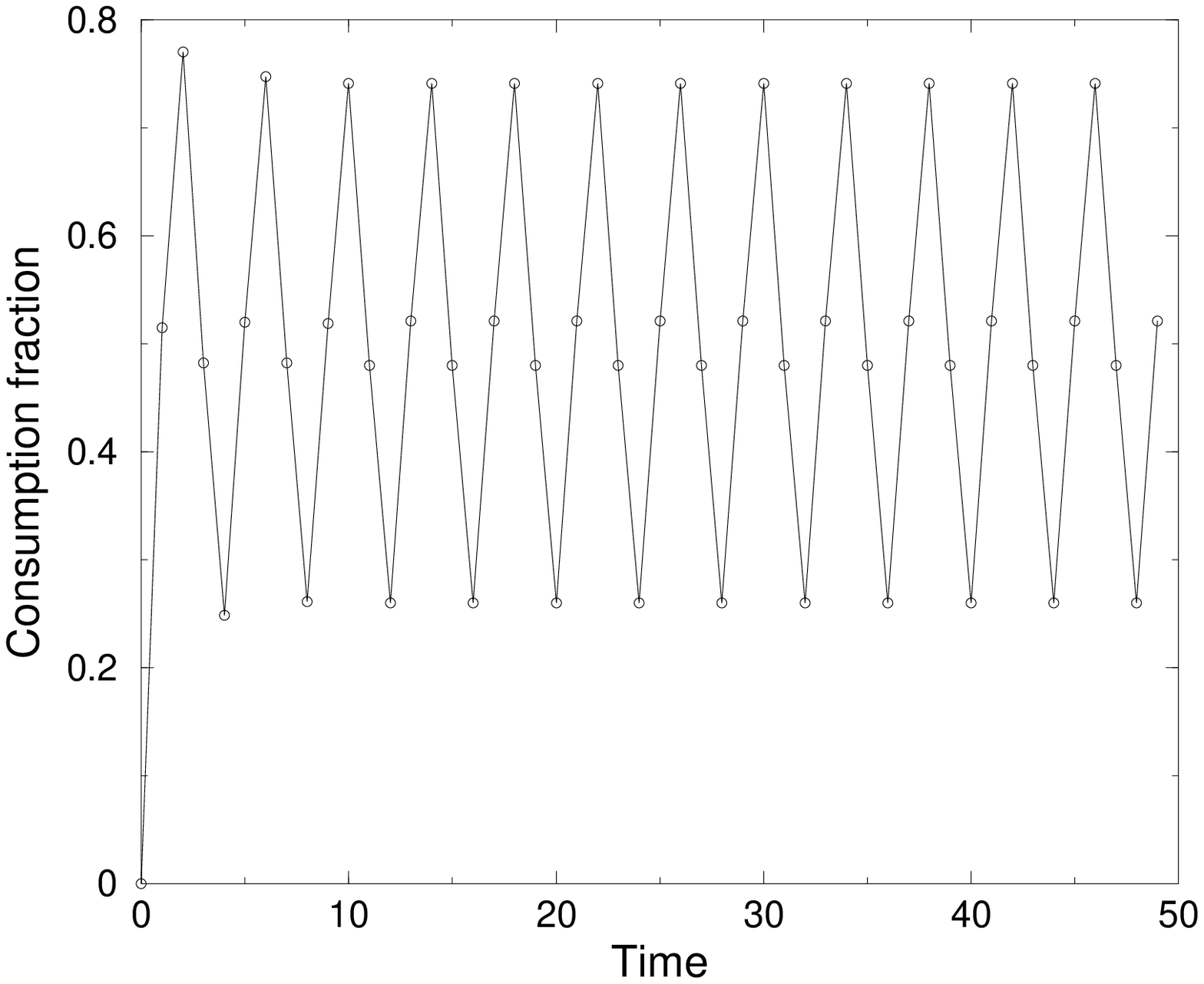}   
\hskip .5cm  \epsfxsize 6.5cm {\epsffile{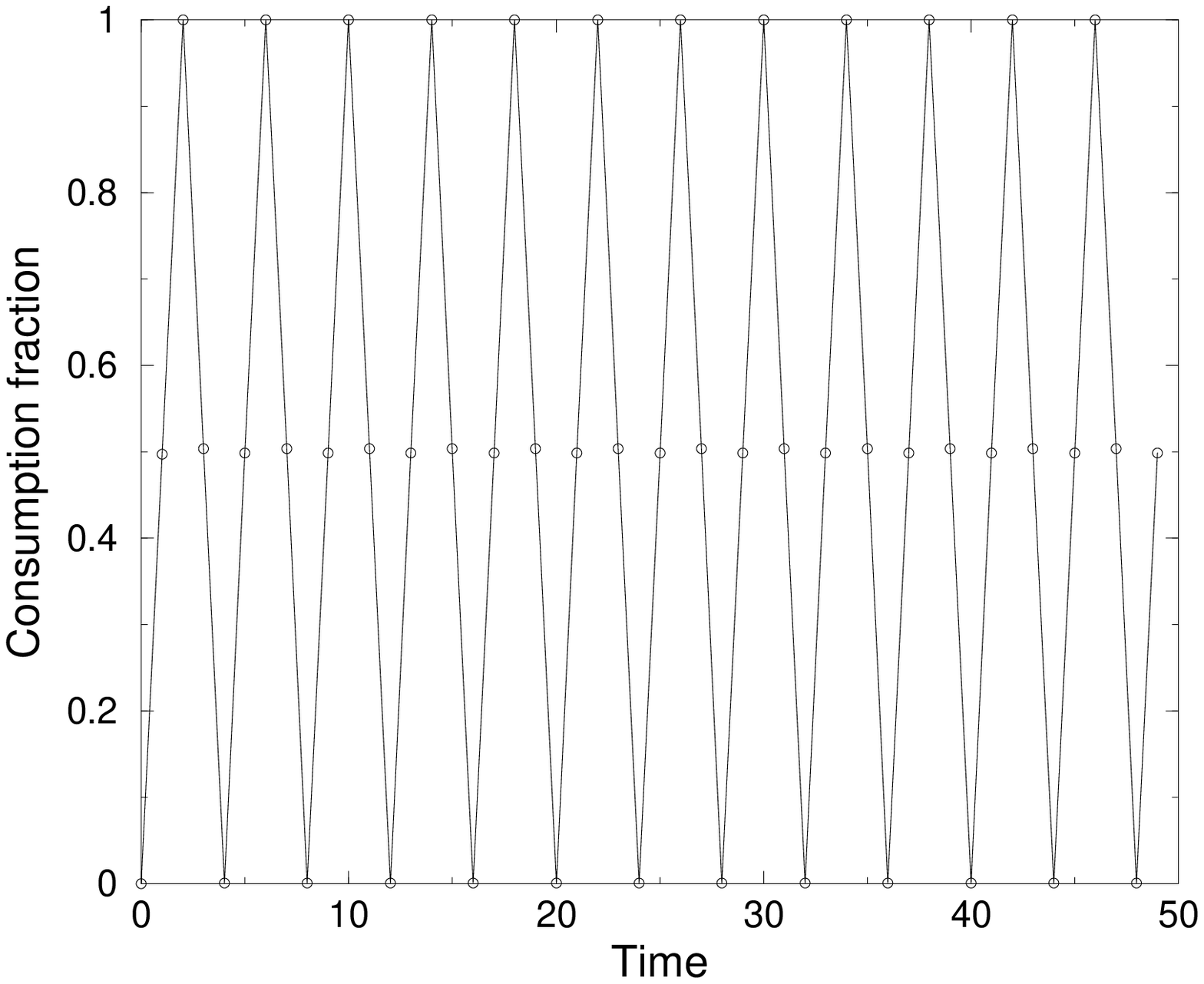}}}
\vskip .5cm
\caption{Total consumption in Model II with  $C=G=N=0$ and   $J_A =
1$ (left). For comparison we show the total consumption in Model I with
$J_A = 1$ and $J_S = 0$ (right).}
 \label{Mod_II_0}
\end{figure}

\begin{figure}[htb]
\vskip 2cm
  \epsfxsize 15cm  \centerline{\epsffile{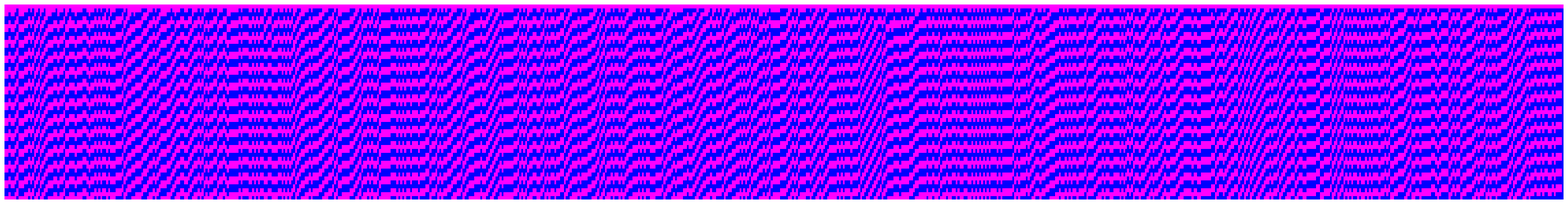}}
\vskip .5cm
  \epsfxsize 15cm  \centerline{\epsffile{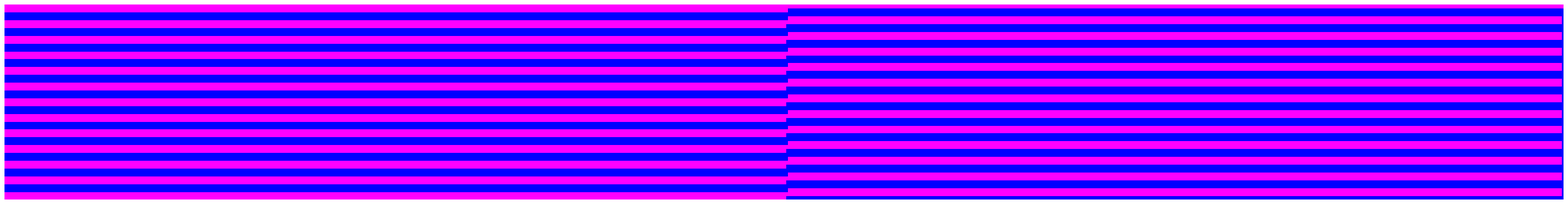}}
\vskip .5cm

\caption{Cycles of consumption in Model II with  $C=G=N=0$ and   $J_A =
1$ (top). For comparison we show the consumption patters in Model I with
$J_A = 1$ and $J_S = 0$ (bottom).}
 \label{Mod_II_1}
\end{figure}

\begin{figure}[htb]
 \begin{center}
  \begin{tabular}[t]{cc}
   \epsfxsize 6cm \epsffile{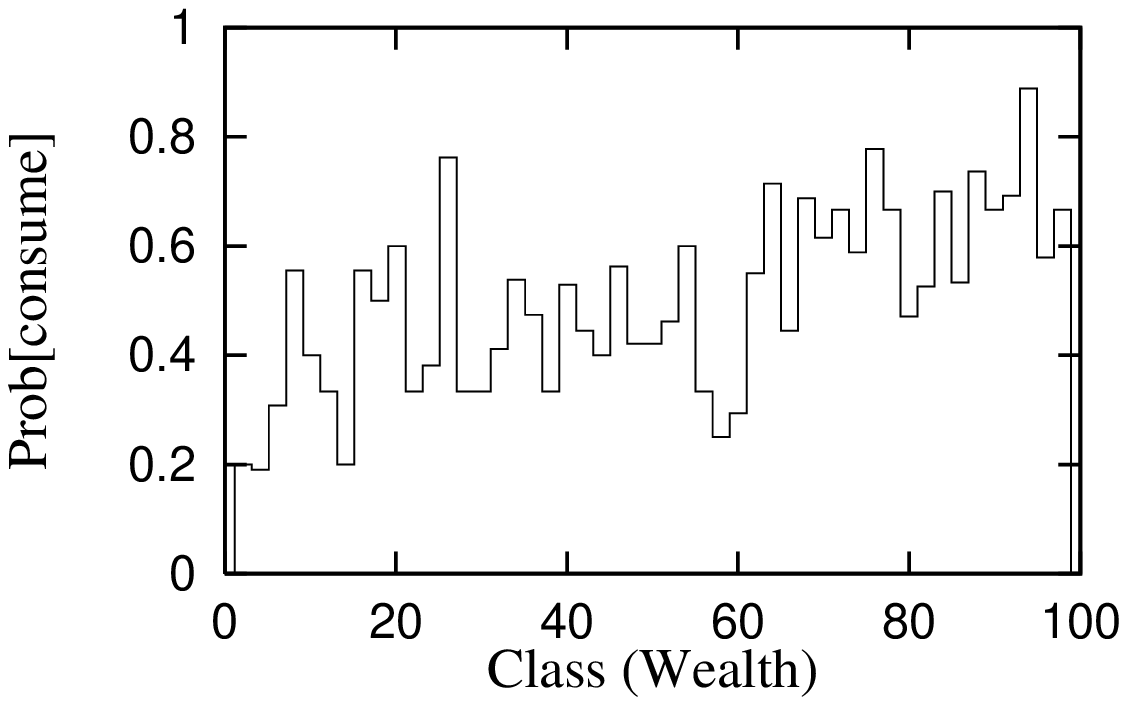} &
   \epsfxsize 6cm \hskip .5cm \epsffile{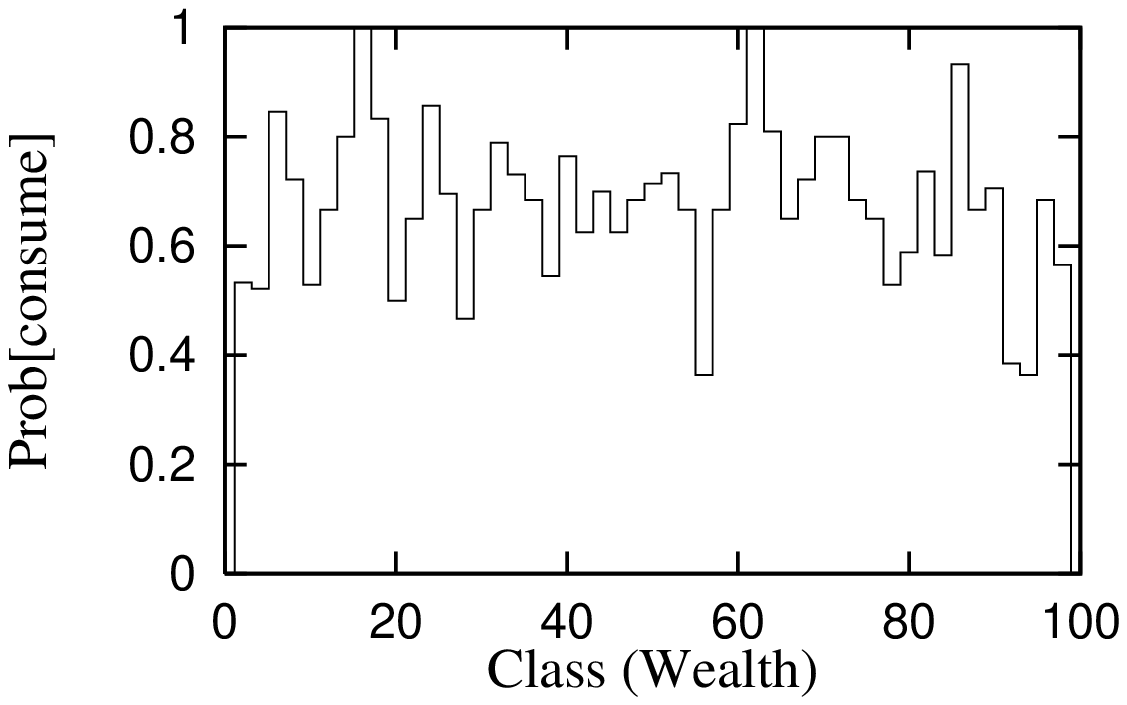} \\
   \epsfxsize 6cm \epsffile{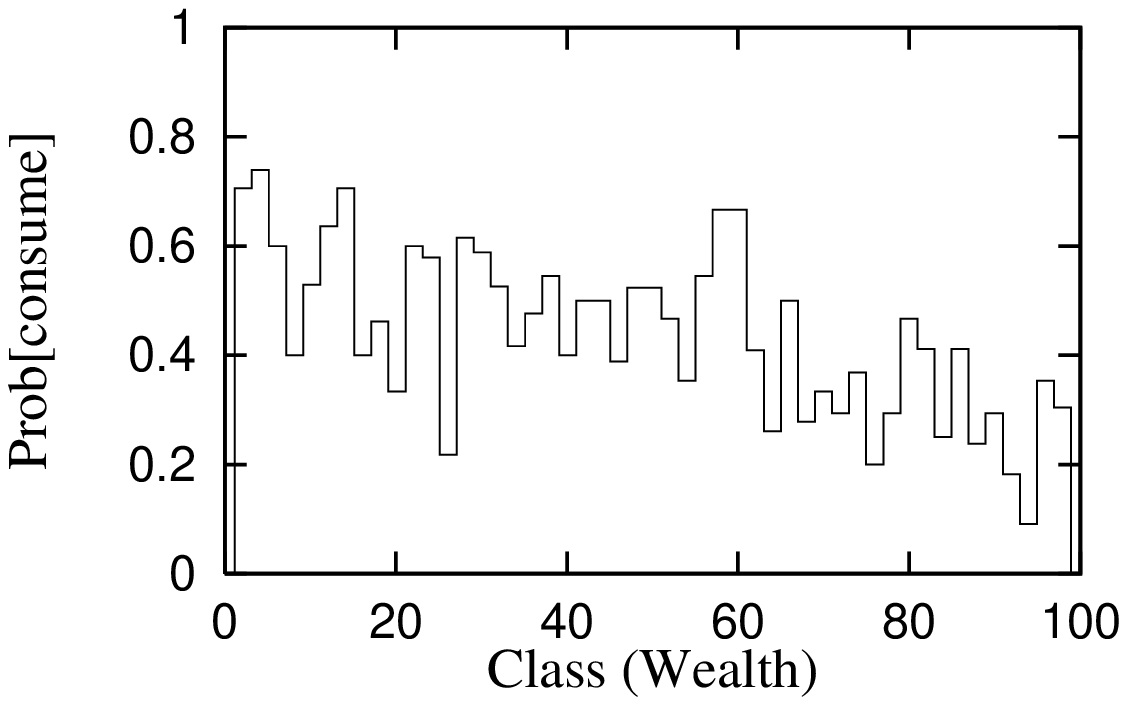} &
   \epsfxsize 6cm \hskip .5cm \epsffile{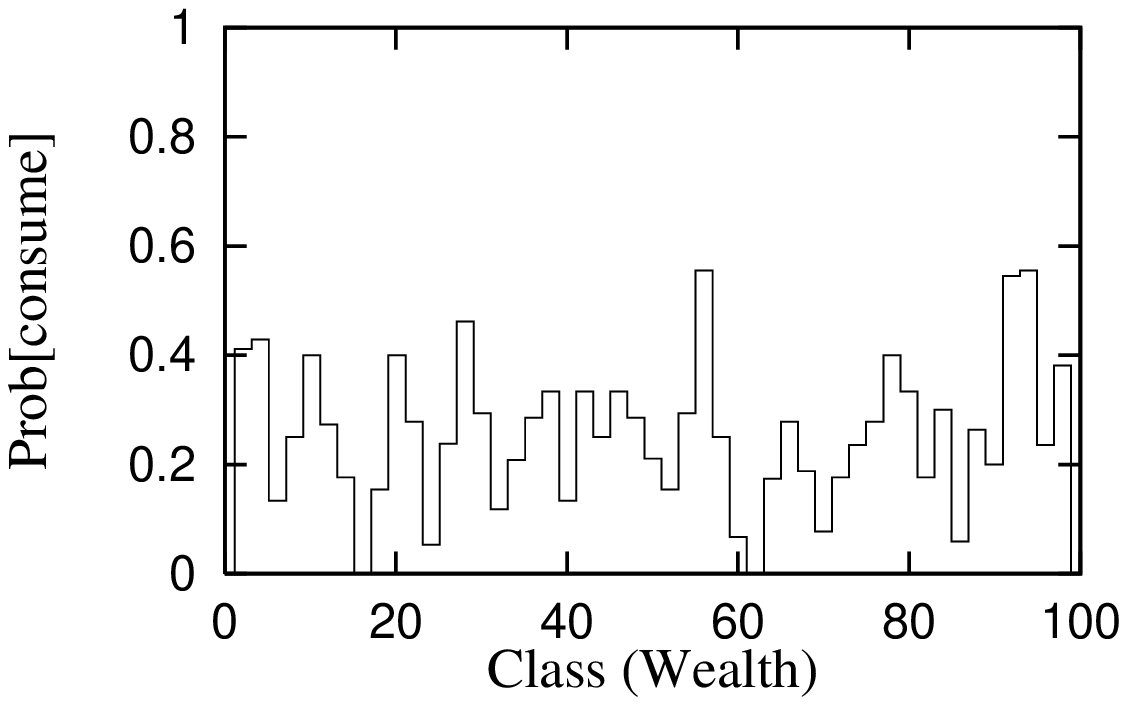}
  \end{tabular}
 \end{center}
\caption{Plots of the distribution of consumption across agents for
Model II at different times (to be read in rows from top left), for
$C=G=N=0$ and $J_A = 1$.} 

 \label{Mod_II_2}
\end{figure}

Starting  with $C=G=N=0$, 
both models show  oscillations in total cosumption,  the amplituide of
which is smaller in Model II compared to Model I (Fig.~(\ref{Mod_II_0})). 
Nevertheless the consumption pattern in the two models is  clearly
different (Fig.~(\ref{Mod_II_1})): in Model II there are no 
diffusing waves  but a more chaotic
situation with different social groups  consuming at different times.
This effect  is better captured by  Fig.~(\ref{Mod_II_2})
where we plot the  consumption density at different locations  along the social
spectrum and at different times. We can see that consumption as a
function of class is multi-peaked.
We also found that the period of oscillations in Model II changes very
little with $J_A$ despite the fact that the potential becomes
longer-range as $J_A$ increases. This is a consequence of  the sharp
increase of the  potential at short distance, so that the main
contribution to the utility function $Y_i$ comes, even at large values of 
$J_A$, mainly from nearest-agent interactions.

\begin{figure}[htb]
\epsfxsize 14cm \centerline{\epsffile{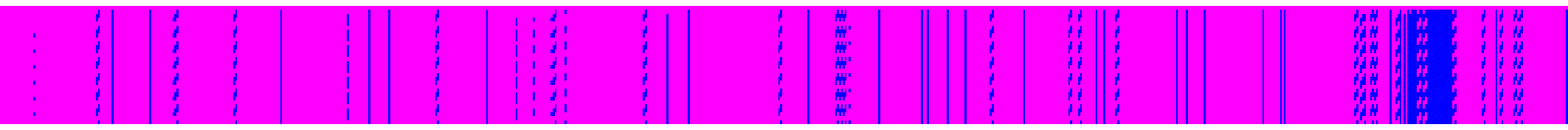}}
\vskip .5cm
\epsfxsize 14cm \centerline{\epsffile{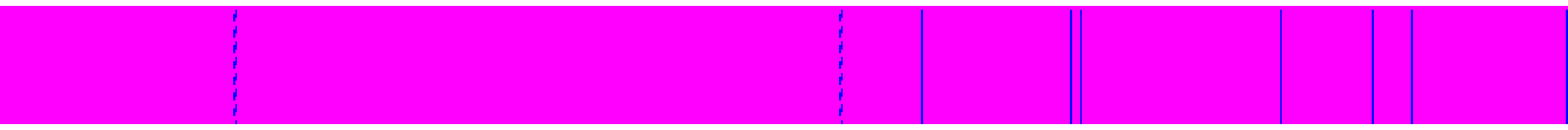}}
\vskip .5cm
\epsfxsize 14cm \centerline{\epsffile{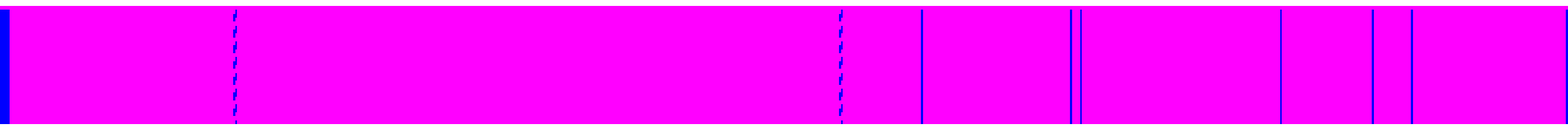}}
\vskip .5cm
\epsfxsize 14cm \centerline{\epsffile{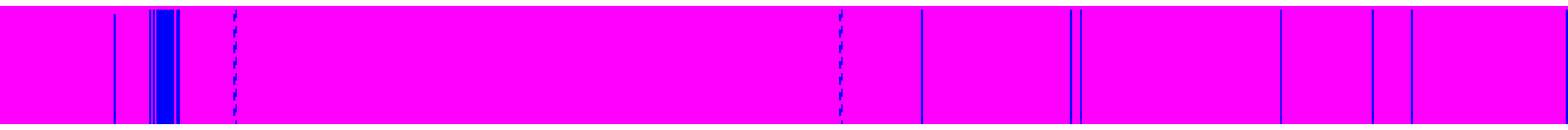}}
\vskip .5cm
\epsfxsize 14cm \centerline{\epsffile{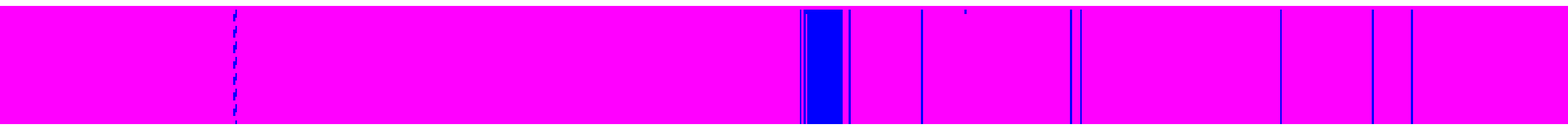}}
\vskip .5cm
\epsfxsize 14cm \centerline{\epsffile{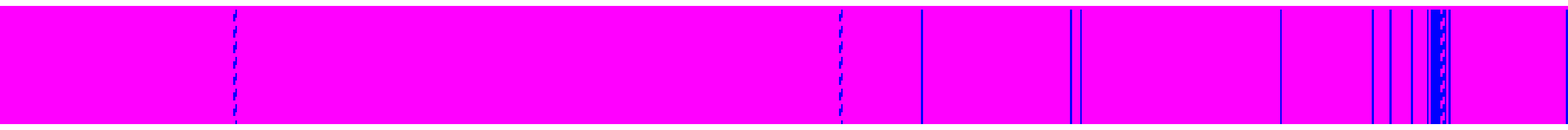}}
\vskip .5cm
\caption{Consumption behaviour in Model II for (from top to bottom) 
(a)$C=0.5, G=0$
(b)$C=4, G=1, w_m = 0 $
(c)$C=4, G=1, w_m = 10 $
(d)$C=4, G=1, w_m = 50 $
(e)$C=4, G=1, w_m = 90 $.}
 \label{Mod_II_costs}
\end{figure}

Adding costs, while keeping $G=N=0$, generates market segmentation.
The larger the costs are (Fig.~(\ref{Mod_II_costs} (a) and (b) ))
the more sparse the consumption becomes. It is interesting though
that even with high costs some agents still choose to consume out of
their desire to distinguish themselves from their peers.

Finally, we study the case of a value good (Fig.~(\ref{Mod_II_costs}
(c), (d) and (e))) which would mostly suit consumers whose  wealth
is close to a reference level $w_m$. In this case  the good finds a stable niche:
those agents who benefit from  its  consumption continue to consume the
good (inducing possibly fringes to their neighours), but their behaviour
does not affect distant social groups (not even when it is for the
richest that the good has an high intrinsic value). 

Adding idiosyncratic preferences in all of the above cases only
generates occasional additional consumption in  the  patterns above
but does not change the asymptotic behaviour of the system.
Let us finally comment that even if we take $J_A$ large (in which case
the interaction becomes long-range), we do not obtain waves that spread
smoothly like in Model I, and this is due to the absense of peering for
nearby consumers.

\section{Conclusions}

In this paper we have focused on a potentially important mechanism that
drives consumption decision: the interaction among heterogeneous
consumers.
In the sociology literature, interactions among individuals, belonging
to similar or different social circles, are often seen as a major
mechanism  that determines new styles of behaviour.
In  model I  we studied how peering, distinction and aspiration effects,
together with  costs and  intrinsic values of a good, generate
different  consumption patterns, under the assumption that  information
on the consumption behaviour of  agents is public ($J_A$ and $J_S$ are
long  range and saturate to a finite value at large distances).
However collective behaviour may be affected by the structure of the
communication channels. 
Models with imperfect information diffusion and social segregation in
the way knowledge is transmitted have  been proposed to explain fashion
cycles (Corneo et al.  1994) . 
We take into account the possible  local nature of interactions in
model II  where the $J_{ij}$ decay rapidily to zero with  social distance.

Both models present different consumption regimes: a dynamical one
characterized by waves and cycles plus a variety of stationary patterns.
The two models, though, appear very different in the way consumption is
spatially distributed.
Initial conditions combined with random individual  preferences
may have a major role in pushing the dynamics of the system into one of the
different  asymptotic states. 

To conclude we point that it would be interesting to compare  patterns of consumptions for
different choices of the   distribution of agents'  wealth, such as, for example, a
discontinuous distribution with gaps in the social spectrum or a
gaussian distribution centered around a given social class. 

\section*{Acknowledgments}
We are grateful to S. Jafarey for helpful comments.
V.K. acknowledges the support of the General Secretariat of Research and
Technology, Greece, under contract 45890. V.K also wishes to thank
the University of Essex for the kind hospitality provided during the
initial stages of this work.

\section*{References}

Aoki, M. {\em New approaches to macroeconomic modeling},
Cambridge University Press (1996). 
\\
\\
 Arenas A.,  Diaz-Guilera A.,  Perez C. J. and Vega-Redondo F., {\it
``Self-organized evolution in socio-economics environments''}, {\tt
cond-mat/9905387} (1999).
\\
\\
  Arthur W. B.,
 Anderson P.W. ,Arrow K. J. and Pines D., eds., The economy as an evolving complex system I, 
Addison-Wesley (1988).
\\
\\
 Arthur W.B.,
Durlauf S.N. and Lane D., eds.,
 The economy as an evolving complex system II, Redwood City:
Addison-Wesley (1997). 
\\
\\
Corneo G. and  Jeanne O.
{\em A Theory of Fashion Based on Segmented Communication}, e-preprint
http://netec.wustl.edu/WoPEc/data/Papers/bonbonsfa462.html (1994).
\\
\\
Cowan~R.,~Cowan~W.~and~Swan~P.,
 {\em Waves in
Consumption with Interdependence among Consumers}, 

\noindent
e-preprint 
http://netec.wustl.edu/WoPEc/data/Papers/dgrumamer1998011.html (1998).
\\
\\
  Crisanti A. and  Sompolinsky H., {\em Dynamics of spin
systems with randomly asymmetric bonds}, Phys. Rev. A {\bf 37}
(1988) 4865.
\\
\\
  Cugliandolo L.F. et al.,
{\em Glassy behaviour in disordered systems with non-relaxational dynamics}, Phys. Rev. Lett. {\bf 78},
350 (1997). 
\\
\\
 Durlauf S., {\em  Statistical mechanics approaches to socioeconomic behavior},
in The economy as an evolving complex system II, W.B. Arthur,
S.N. Durlauf and D. Lane, eds., Redwood City: Addison-Wesley (1997).
\\
\\
Iori G. and   Marinari E., 
{\em On the Stability of the Mean-Field Glass Broken Phase under
Non-Hamiltonian Perturbations}
 J. Phys. A: Math. Gen.  (1997)
4489-4511. 
\\
\\
Kirman, A. {\em Economies with interacting agents},
in The economy as an evolving complex system II, W.B. Arthur,
S.N. Durlauf and D. Lane, eds., Redwood City: Addison-Wesley (1997).
\\
\\
 Mezard M. et al., {\em ``Spin Glass theory and Beyond''},
(World Scientific, Singapore 1987).
\\
\\
 Persons J. and  Warther V. {\em Boom and Bust Patterns in the Adotpion of
Financial Innovations} e-preprint 
http://netec.wustl.edu/WoPEc/data/Papers/wopohsrfe9601.html (1995).

\vskip 2cm

 \newpage

\vskip 1cm

\newpage

\newpage

\end{document}